\documentclass[%
 reprint,
 amsmath,amssymb,
 aps,
]{revtex4-1}

\usepackage{graphicx}
\usepackage{dcolumn}
\usepackage{bm}
\usepackage{lipsum}

\usepackage{subfigure}

\begin{document}


\title{
Survival chances of a prey swarm:\\ 
how the cooperative interaction range affects the outcome\\
}

\author{Dipanjan Chakraborty, Sanchayan Bhunia, Rumi De}
\email{Corresponding author:rumi.de@iiserkol.ac.in}
\affiliation{%
 Indian Institute of Science Education and Research Kolkata, Mohanpur, Nadia-741246, India\\
 }
%
%
%


\begin{abstract}
A swarm of preys when attacked by a predator is known to rely on their cooperative interactions to escape. 
Understanding such interactions of collectively moving preys and the emerging patterns of their escape trajectories still remain elusive.
In this paper, we investigate  how the range of cooperative interactions within a prey group affects the survival chances of the group while chased by a predator. As observed in nature, the interaction range of preys may vary due  to  their vision, age, or even physical structure. Based on a simple theoretical prey-predator model, here, we show that an optimality criterion for the survival can be established on the interaction range of preys. Very short range or long range interactions are shown to be inefficient for the escape mechanism. Interestingly, for an intermediate range of interaction, survival probability of the prey group is found to be maximum. Our analysis also shows that the nature of the escape trajectories strongly depends on the range of interactions between preys and corroborates with the naturally observed escape patterns. Moreover, we find that the optimal survival regime depends on the prey group size and also on the predator strength.
     
\end{abstract}

\pacs{Valid PACS appear here}
\maketitle


\section*{Introduction}
In nature, cohesive group formation has been observed in diverse species  ranging from bacterial colony to flocking of birds, swarming of insects, schooling of fishes, huddling of penguins  to name a few \cite{vicsekreview2012, cavagnareview2018,  Kparrishscience, krausebook2002, sumpterbook2010, ballerini2008, Arnoldplosone2012, jmichaelscience2009, abaid2010interface, couzin2003id}.
Swarm behaviour arises due to variety of reasons as individual members gain mutual benefits from one another belonging in a group while searching for food, finding new nests, migrating from one place to another, or to overcome various environmental hurdles in general
\cite{vicsekreview2012, cavagnareview2018, Arnoldplosone2012, traniello2003ARE}.
Another major factor of forming group is thought to be due to predation avoidance where survival chances in a group turns out to be better than
solitary individuals \cite{Kparrishscience,penzhornbook1984,pitcherbook1983}. 
Moving in a large group often  dilute the encounter and increases the overall alertness since many eyes could keep a careful watch for a possible danger or a predator attack. It also confuses the predator by making it difficult to focus on any particular member among a large group 
of preys \cite{neilletal}. However, cohesive movements could also be unfavourable for preys as the predator can then easily track the group and attack them. 
For example, fish schools are easily tracked and caught by marine predators \cite{parrishenvbio}.
Moreover,  preys at the boundaries and the trailing ones are more vulnerable for predation, so preys compete within for the protected position. Competitions may also arise due to limited food resources or due to aggression within the group.
Thus, there found to be  often a trade off between staying together versus individual needs.
So, prey groups always look for efficient strategies for the survival from predator attacks \cite{kerleyjoz2005, Mckenzieinterfacefocus2012, carobook2005,Humpherisoecologia1970}. 
There are different escape strategies have been observed in nature.  
For example, a school of marine fish  would scatter fixing the predator at the centre or splitting up into subgroups creating visual confusion \cite{pitcherbook1983,Partridge1982}.
Besides, on finding a potential threat, animal aggregation often  moves closer to reduce the chance of being caught by the predator \cite{hamiltonjtb}.
Moreover, there are instances of direct escapes where preys simply straight away run in the opposite direction to escape from the predator, or run in random zigzag motion to confuse the predator.
It is further observed that preys often interact within the group to avoid predation by opting different kind of swarming  patterns like spinning, circling,
splitting up into sub groups etc \cite{carobook2005, Humpherisoecologia1970,DomeniciJEB2011,EdutBBR2004,DomeniciMB1997}.
However, it still remains far from clear how the local interactions among swarming preys lead to complex behavioural patterns,  or how preys optimize their  survival chances or influence the predation rates etc.

There are several experimental and theoretical studies which have contributed immensely to understand the  emergent behaviours of swarming in living organims \cite{vicsekswarm2008,attanasi2014collective,calovietal,couzinetal2005,couzinetal2002,LeePRE2015}.
Considerable efforts have also been made to understand the collective dynamics of prey-predator systems. 
Detailed studies on escape trajectories of different species under threats show a certain degree of unpredictability in  their escape patterns that confuse the predator in the chase \cite{DomeniciJEB2011, DomeniciMB1997}.
Besides, how the size of the prey group affects predator attacks and the success rate have also been investigated in the field \cite{cresswell2011predicting}. 
Also, it has been observed that cooperativity in predator groups significantly increase hunting success upto a certain threshold size of the group \cite{schallerbook1972, boeschAOPA1989, geseCJZ2001}.
However, understanding local interactions within the prey group in natural field is quite challenging due to unpredictable nature of the predator attack. In such scenarios, theoretical models further help us to get insights into the complex dynamics of collectively interacting systems. 
For example, based on self-propelled particle models, collective predation and escape strategies have been explored to provide insights into the predation rate and the catch time of the group \cite{AngelaniPRL2012}. 
In another simple model of prey-predator system, it has been shown that prey swarm could easily escape from the weaker predator but as the strength of the predator increases, system passes through a transition from  confusion state of the predator to chasing dynamics \cite{chenJRSCinterface2014}.
There are also other models on swarming behaviour of preys in the presence of predators where different force laws between predators are 
explored \cite{ZhdankinPRE2010}. 
Predator confusion and its effect on reducing the attack  to kill 
ratio has also been studied \cite{krausebook2002,bazzietal}.
Another evolutionary model suggests that predator confusion drives the swarming behaviour of preys and the attack efficiency decreases rapidly when predators visual filed is restricted  \cite{OlsonJRSC2013}.
Further, survival of a prey has been studied by assigning different sighting radius to the prey and its predators on a square lattice that suggests the importance of optimal sighting range for effective 
evasion \cite{Oshaninproceedings2009}.

Indeed, in natural scenario, the range of interactions of preys  may be limited due  to  their sensitivity, vision, age, or even physical structure.
However, very little has been known about the range of cooperative interactions among  preys under a predator attack.
It is observed that the prey groups rely on their local interactions to confuse the predator. 
Coordinating the movement of individuals in a group is important to ensure an escape.
In this paper, we investigate the effect of range of cooperative interactions among preys in a group while chased by a predator. Based on a simple theoretical prey-predator model  that incorporates the essential interactions  between preys and the predator,
we study the escape dynamics and the survival probability of the prey group by varying the interaction range among preys under a predator attack.
Our analysis shows that the range of cooperative interaction has a strong influence on the escape trajectories of preys. It also hugely alters the survival outcome of the prey group. 
Cohesive interactions with the entire group or no interactions among preys appear to be unfavourable for their survival. 
Interestingly, we find that the survival of the group is maximum within an intermediate ranges of interaction radius. The optimal regime further varies depending on the size of the prey group and on the strength of the predator. In addition, we also analyse how  the spatial correlations among preys and the collective ordering of the group get affected with the change in interacting radius. 

\section*{Theoretical Model}
  \begin{figure}[!t]
	\centering
	\includegraphics[width=6.5cm,height=5.5cm]{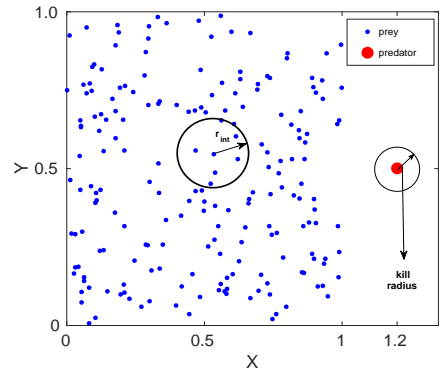}
	\caption{An illustration of initial configuration of the prey group and the predator. The smaller dots (blue) represent the position of preys and the bigger dot (red) is the position of the predator. Here, $r_{int}$ denotes the interaction radius of each prey in the group. The kill radius is shown by the circle around the predator. (color online)}
	\label{fig:Initialconfig}
\end{figure} 
In our model, we consider a group of $N$ preys represented by active particles on a two dimensional space as illustrated in Fig. \ref{fig:Initialconfig}. Each prey is
characterized by the position, $\Vec{r}_i$, and the velocity, $\Vec{v}_i$.
To mimic the physical scenario in the field, we consider that preys move in open space (unlike other studies where periodic boundary condition has been imposed).
We focus on the escape dynamics when the prey group is under attack by a predator. 
In general, due to physical or sensory constraints, it is not possible for preys to interact with all other preys in a large group at the time of escape \cite{kunzthomas2006}.
Therefore, we consider that each prey interacts with the neighbouring  preys within a certain  reaction radius, $r_{\rm int}$, for their survival.
We model the prey-prey interactions  following the existing literature \cite{chenJRSCinterface2014}.
Each prey interacts with the surrounding preys within the reaction radius by  long range attraction and short range repulsion.
The prey-prey interaction force for the $i$'th prey
is given by averaging over all interacting preys within the reaction radius, $r_{\rm int}$, 
\begin{equation}
		\vec{F}_{i,\rm prey-prey}=
	\frac{1}{N_{\rm int}}
	\sum_{j=1}^{N_{int}}\left({\beta (\vec{r_j} - \vec{r_i}) - \alpha 
		\frac{\vec{r_j}-\vec{r_i}}{|\vec{r_j} -\vec{r_i}|^2} }\right); 
		\nonumber		
	\end{equation}
where $N_{int}$ is the number of preys interacting with the $i$'th prey within the given radius, $r_{\rm int}$. 
Here, $\beta$ denotes the strength the prey-prey attraction and $\alpha$ is the strength of repulsion.
Moreover, as preys always try to escape from the predator, the prey-predator interaction is modelled as a repulsive radial force, 
\begin{equation}
	{\vec{F}_{i, \rm prey-predator}}= - \gamma \frac{\vec{r_p} - \vec{r_i}}{|\vec{r_p} - \vec{r_i}|^2}; \nonumber
\end{equation}
here, $\vec{r}_p$ denotes the position of the predator and  $\gamma$ is the strength of repulsive interaction between the prey and the predator. 
On the other hand, as the predator chases the prey group, it could track all preys and its motion is governed by the attractive force averaged over all preys given by,
\begin{equation}
	{\vec{F}_{ \rm predator-prey}}=\frac{\delta }{N}
	\sum_{i=1}^{N}\frac{\vec{r_i} - \vec{r_p}}{|\vec{r_i} - \vec{r_p}|^3}; \nonumber
\end{equation}
where $\delta$ signifies the strength of the predator. 
Hence, the predator-prey interaction decays as the distance between preys and the predator increases. We assume that when the prey comes close to the predator within a certain kill radius, as illustrates in  
Fig. \ref{fig:Initialconfig}, the prey is killed.
Here, we note that the prey-predation interaction could also be considered by different power laws as discussed by chen and kolkolnikov \cite{chenJRSCinterface2014}. 
Now, the equation of motion of preys and the predator can be described by, 
\begin{eqnarray}
	\mu\frac{d\vec{r}_i}{dt} & = & \vec{F}_{i,\rm prey-prey} + \vec{F}_{i,\rm prey-predator}, \label{preyequation}\\
	\mu_{\rm pd}\frac{d\vec{r}_p}{dt} & = & \vec{F}_{\rm predator-prey}.
	\label{predequation}
\end{eqnarray}
Here, $\mu$ and $\mu_{\rm pd}$ represent the coefficient of the viscous drag experienced by the prey and the predator respectively.
In our model, for simplicity we consider the dynamics in over damped limit.

We study  the prey-predator dynamics in dimensionless units.
Position variables are scaled as, $\vec{R}_i=\vec{r}_i/l_0$, $\vec{R}_p=\vec{r}_p/l_0$ and the interaction radius as, $R_{int}=r_{int}/l_0$. The dimensionless time is defined by $T=t/\tau$; where $l_0$ and 
$\tau$ represent the characteristic length and time scale of the prey system.  Also, the other relevant scaled parameters are given as, $\alpha_0=\alpha/\tau \mu$, $\beta_0=\beta/\tau \mu$, $\gamma_0=\gamma/\tau \mu$, and 
$\delta_0=\delta/l_0 \tau \mu_{\rm pd}$.

\section*{Results}
\begin{figure}[!t]
	\centering
	\subfigure[ ]{\label{fig:a}\includegraphics[width=2.8cm,height=2.8cm]{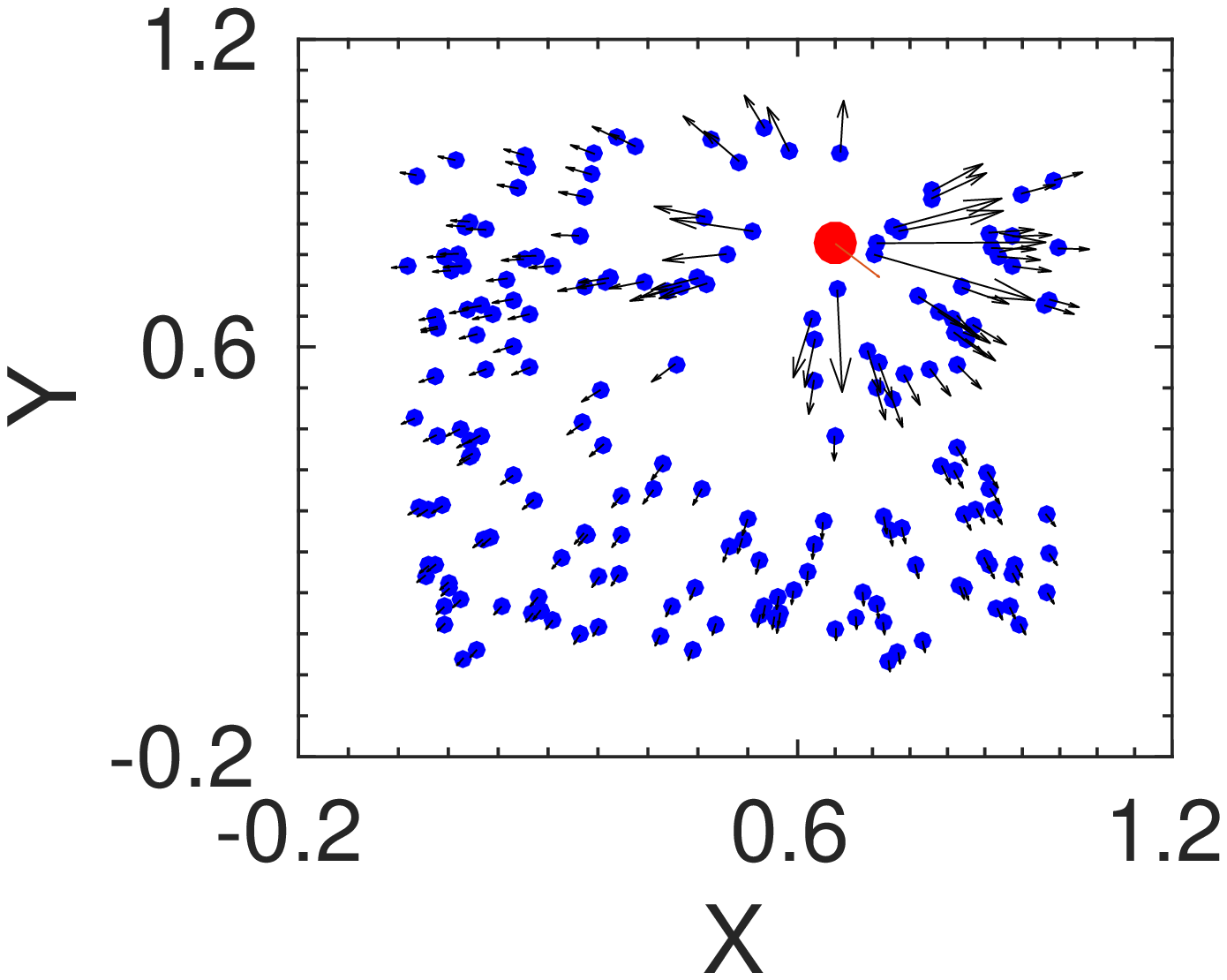}}
	\subfigure[ ]{\label{fig:b}\includegraphics[width=2.8cm,height=2.8cm]{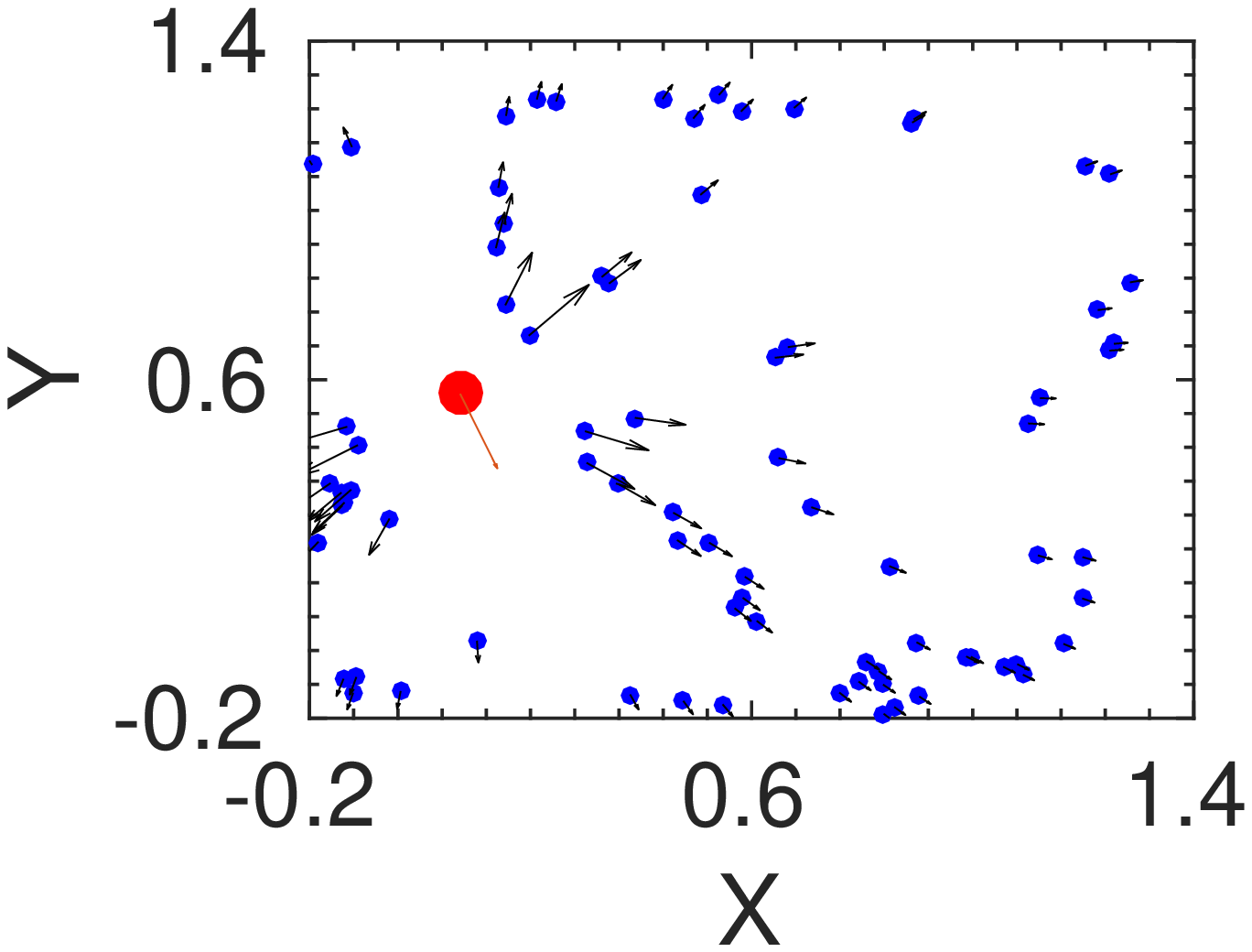}}
	\subfigure[ ]{\label{fig:c}\includegraphics[width=2.8cm,height=2.8cm]{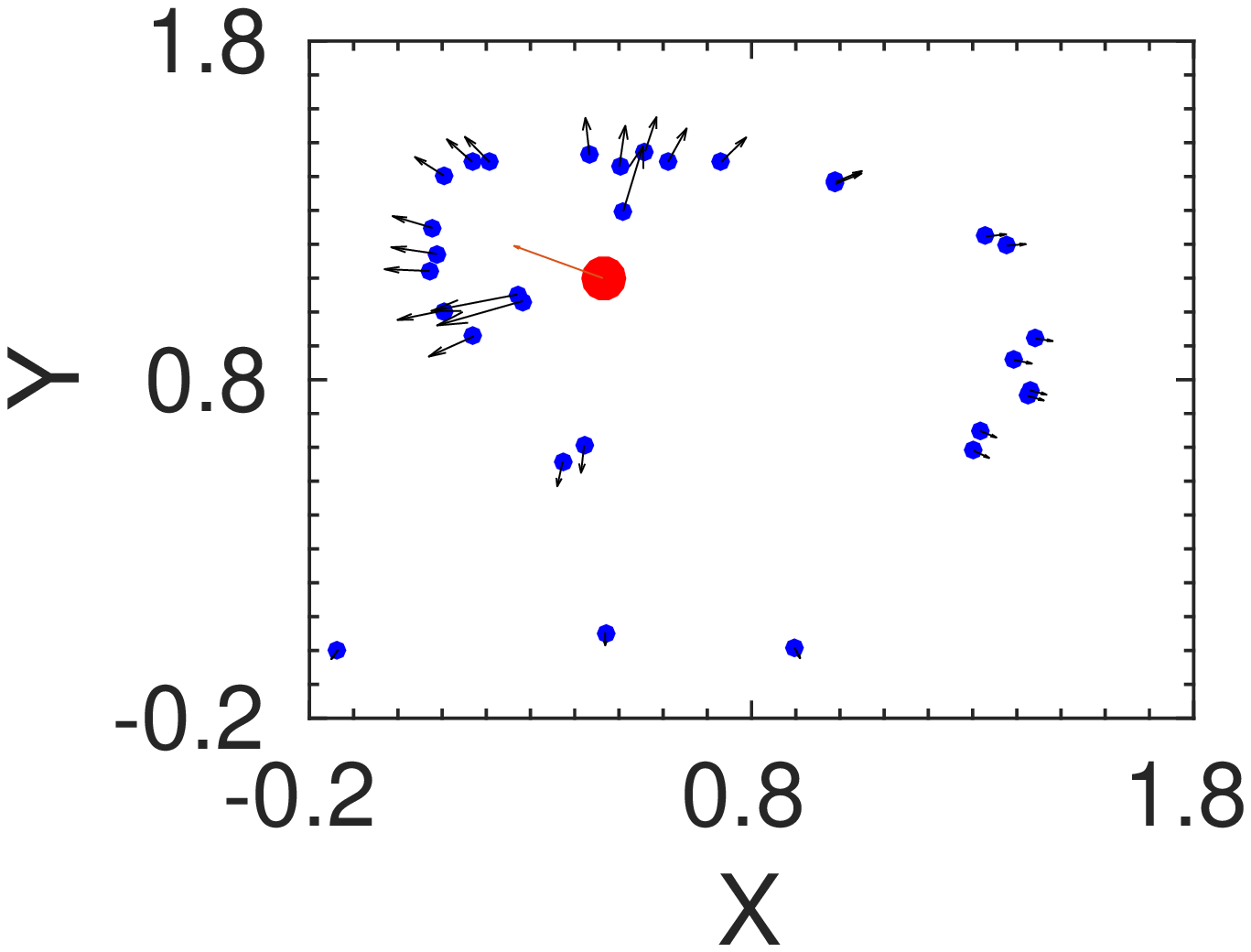}}
	
	\subfigure[ ]{\label{fig:d}\includegraphics[width=2.8cm,height=2.8cm]{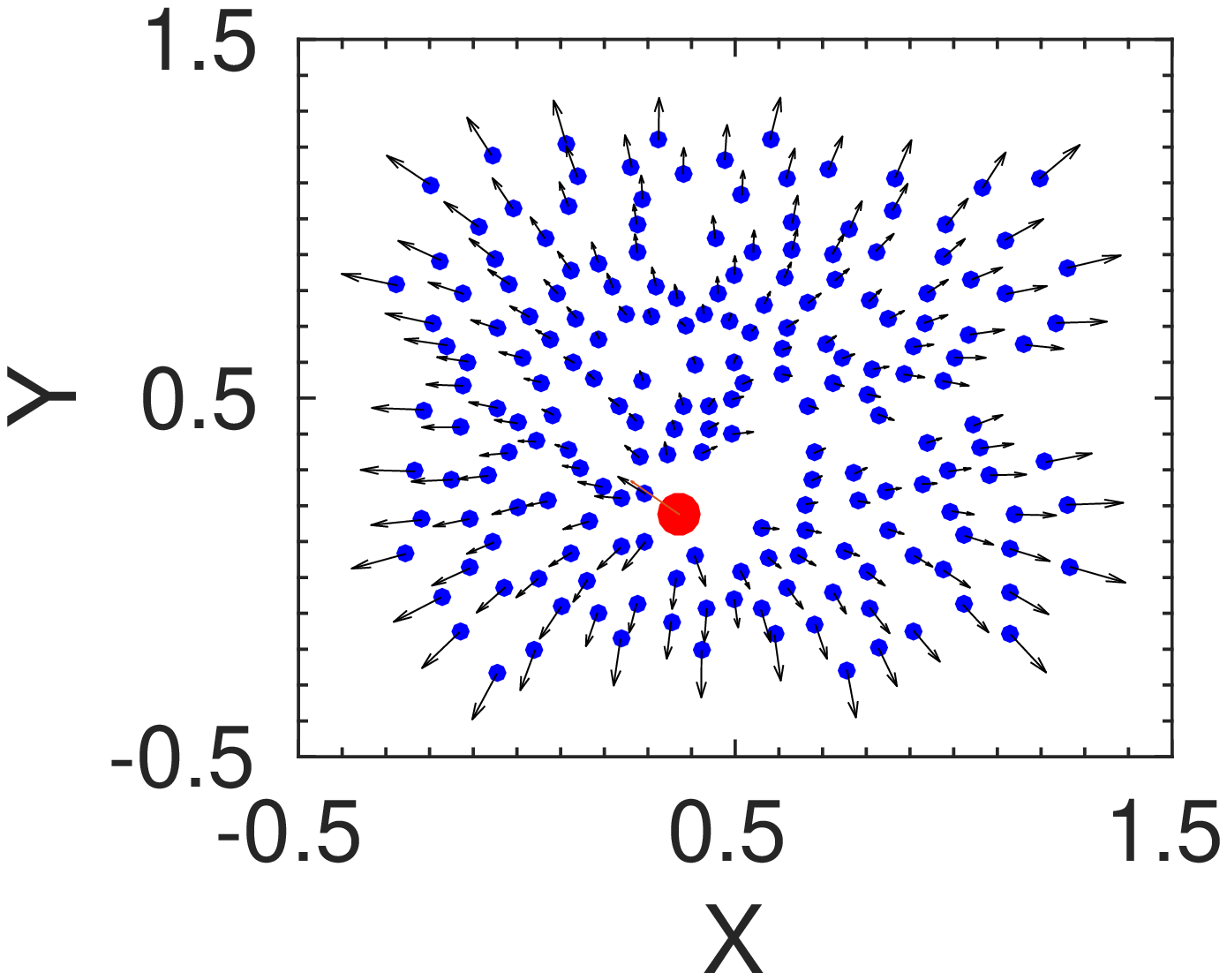}}	
	\subfigure[ ]{\label{fig:e}\includegraphics[width=2.8cm,height=2.8cm]{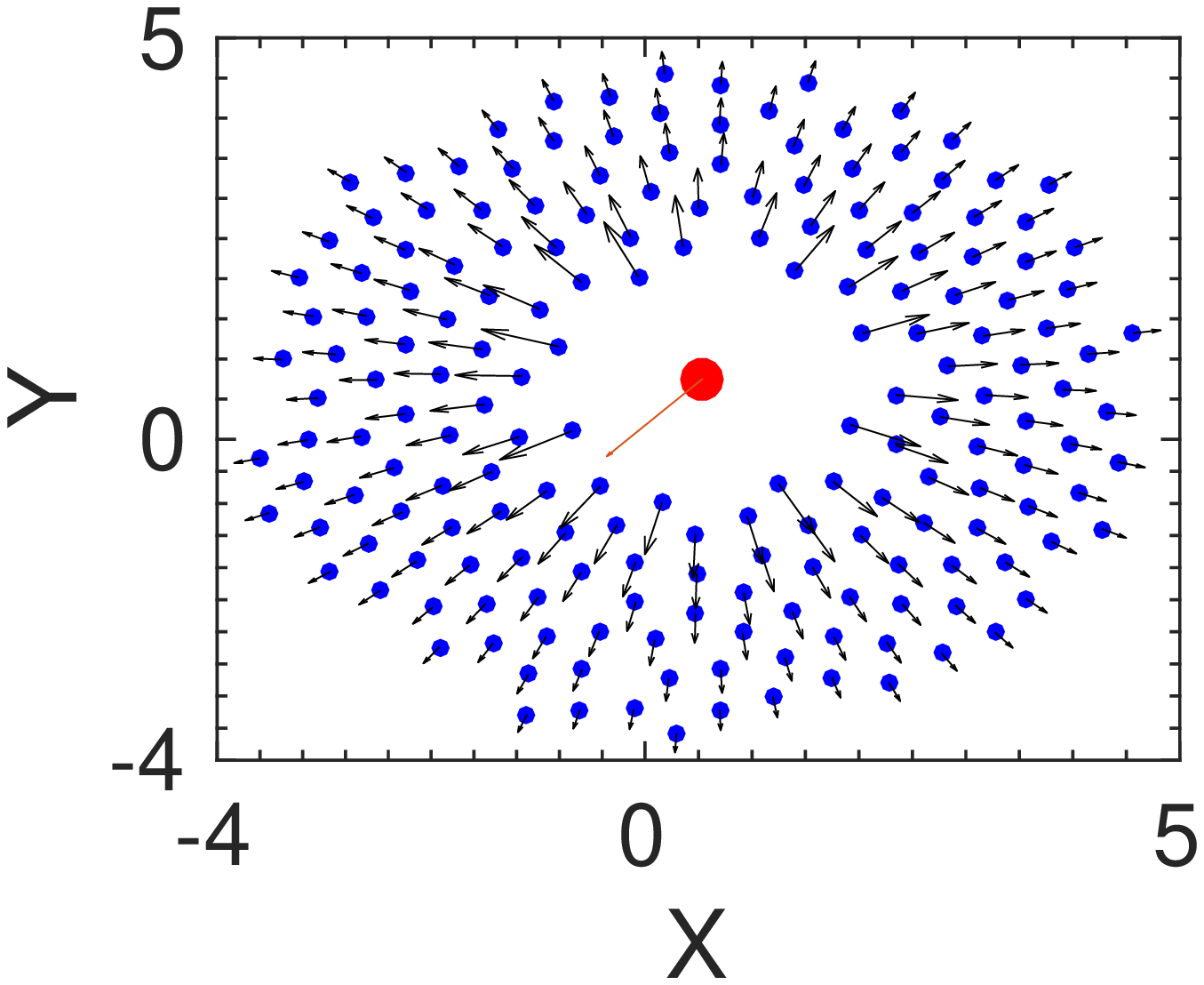}}
	\subfigure[ ]{\label{fig:f}\includegraphics[width=2.8cm,height=2.8cm]{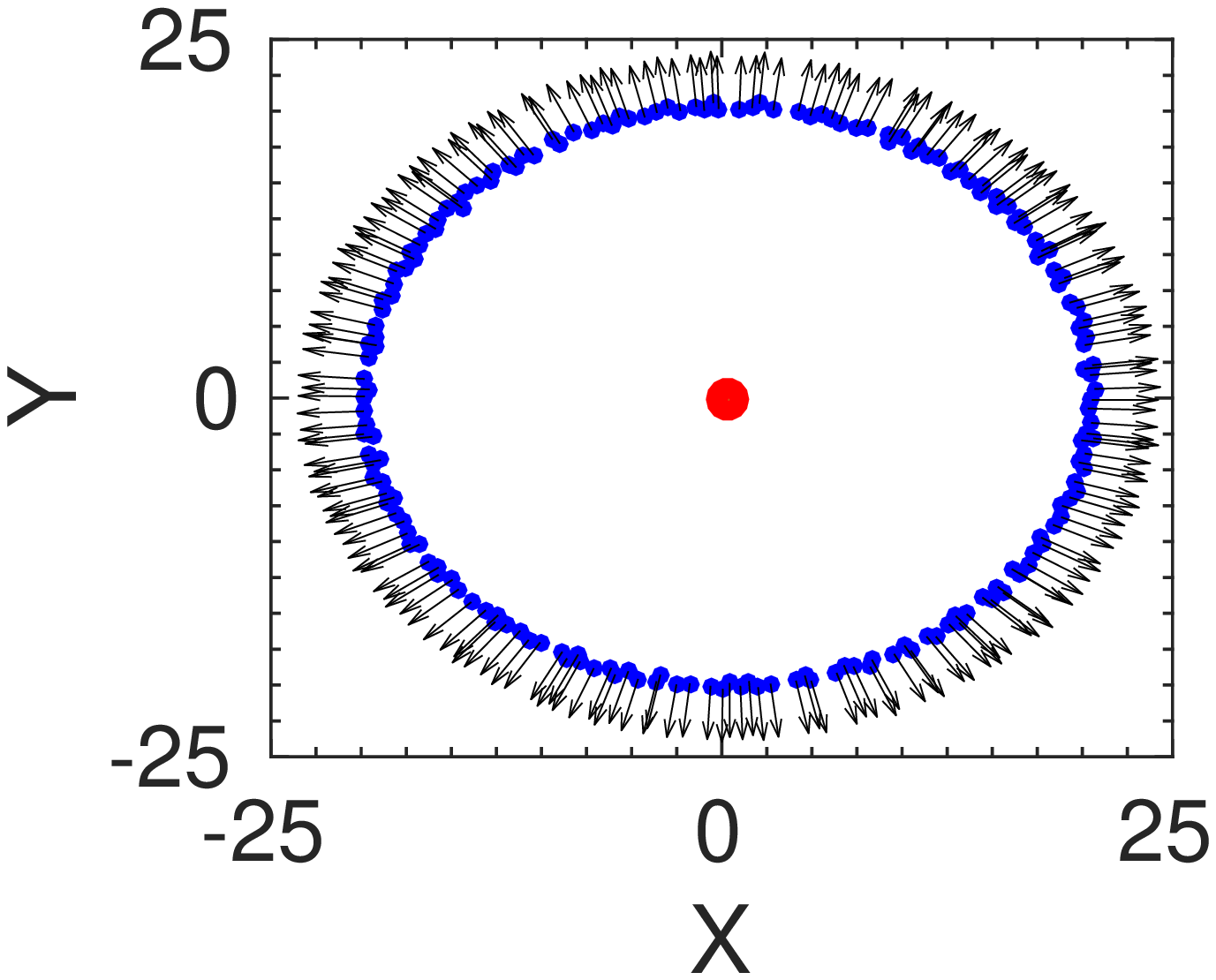}}
	
	\subfigure[ ]{\label{fig:g}\includegraphics[width=2.8cm,height=2.8cm]{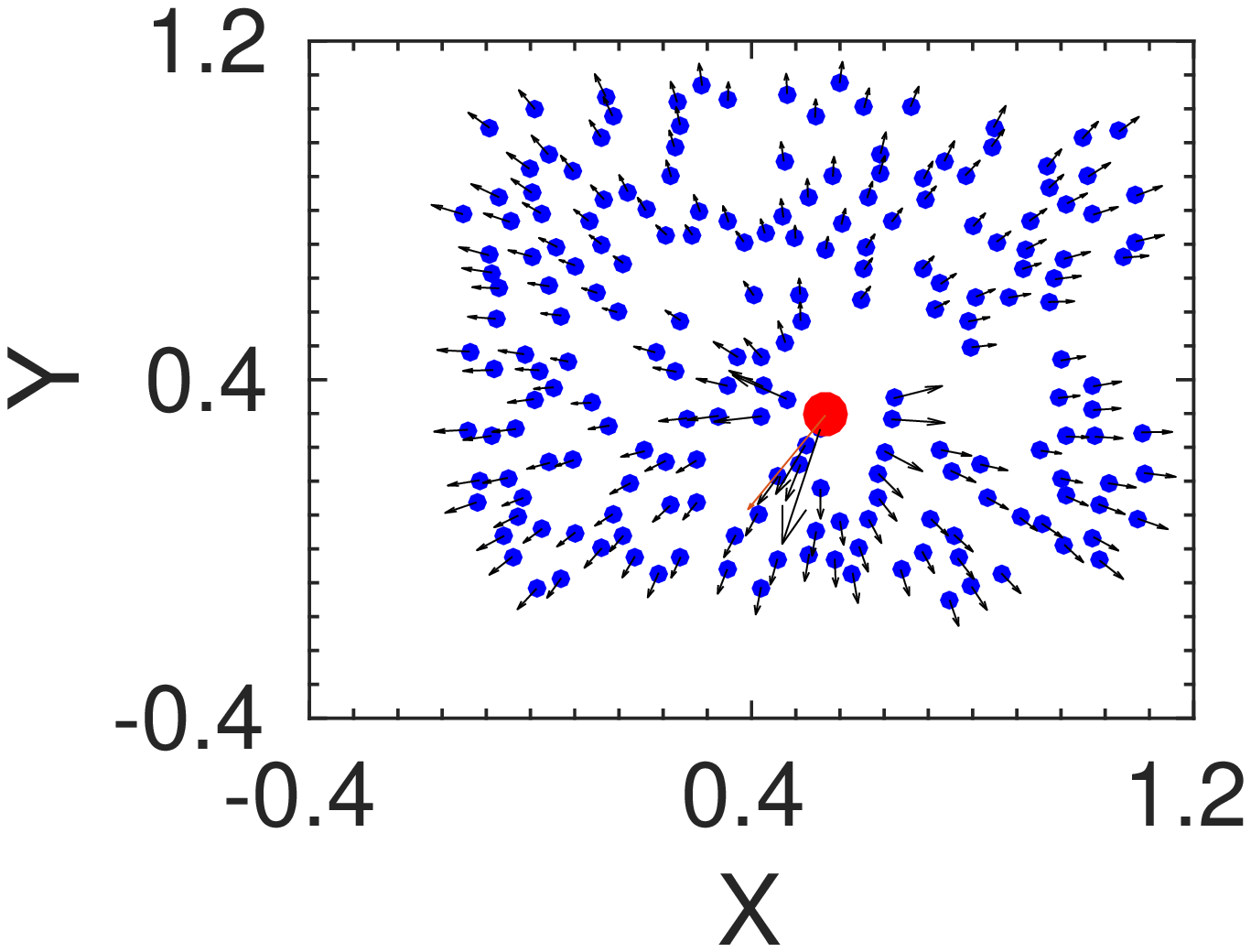}}
	\subfigure[ ]{\label{fig:h}\includegraphics[width=2.8cm,height=2.8cm]{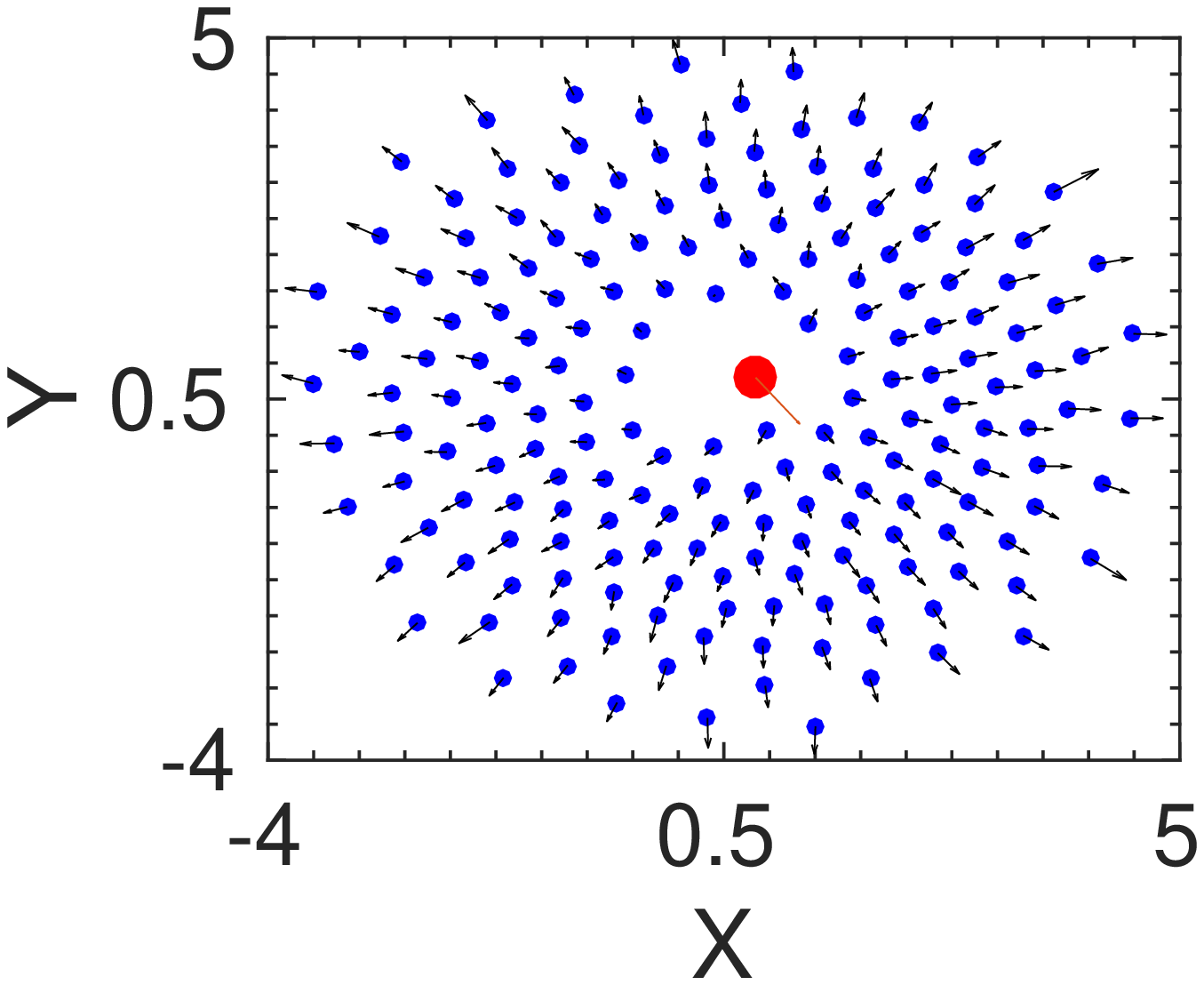}}
	\subfigure[ ]{\label{fig:i}\includegraphics[width=2.8cm,height=2.8cm]{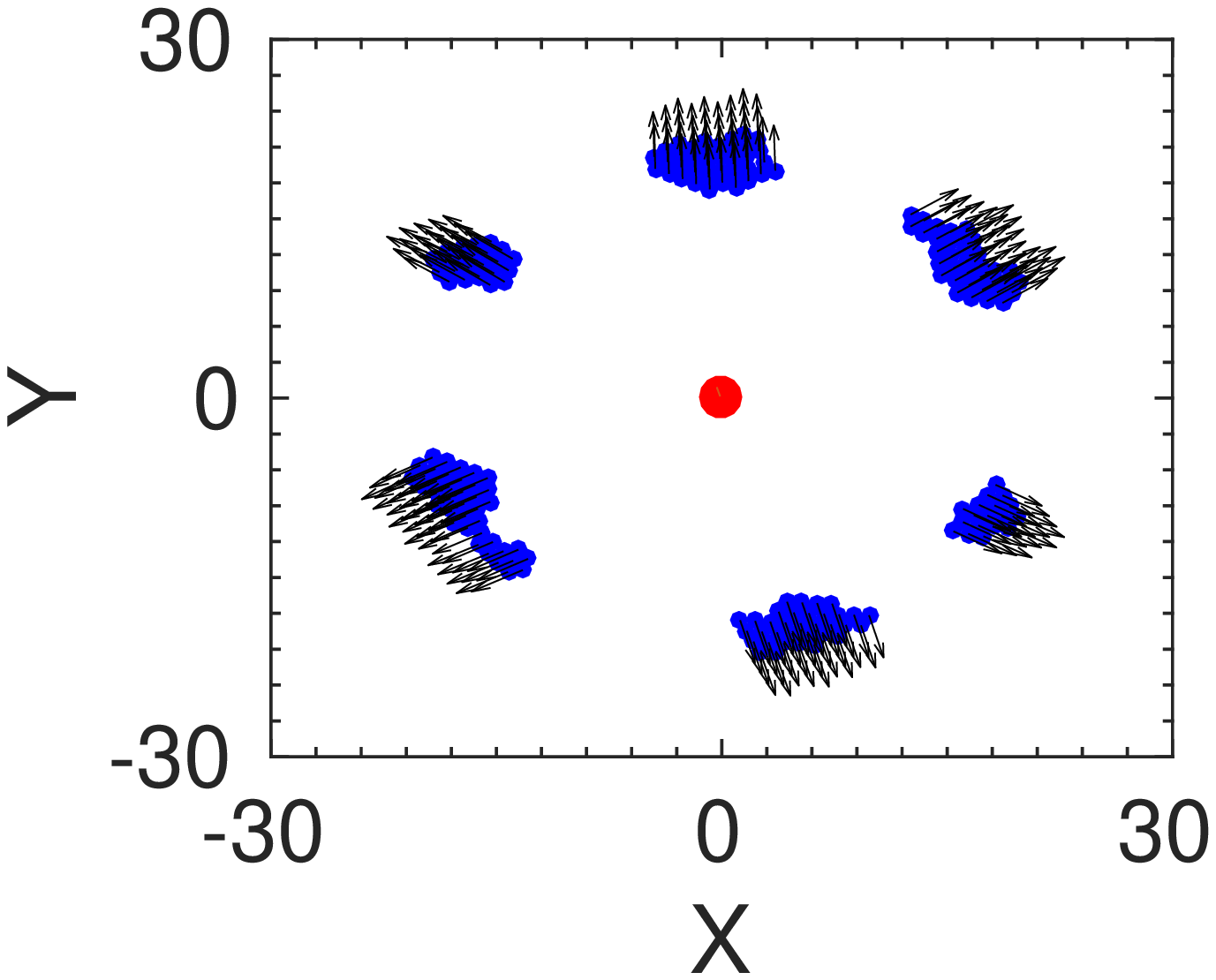}}
	
	\subfigure[ ]{\label{fig:j}\includegraphics[width=2.8cm,height=2.8cm]{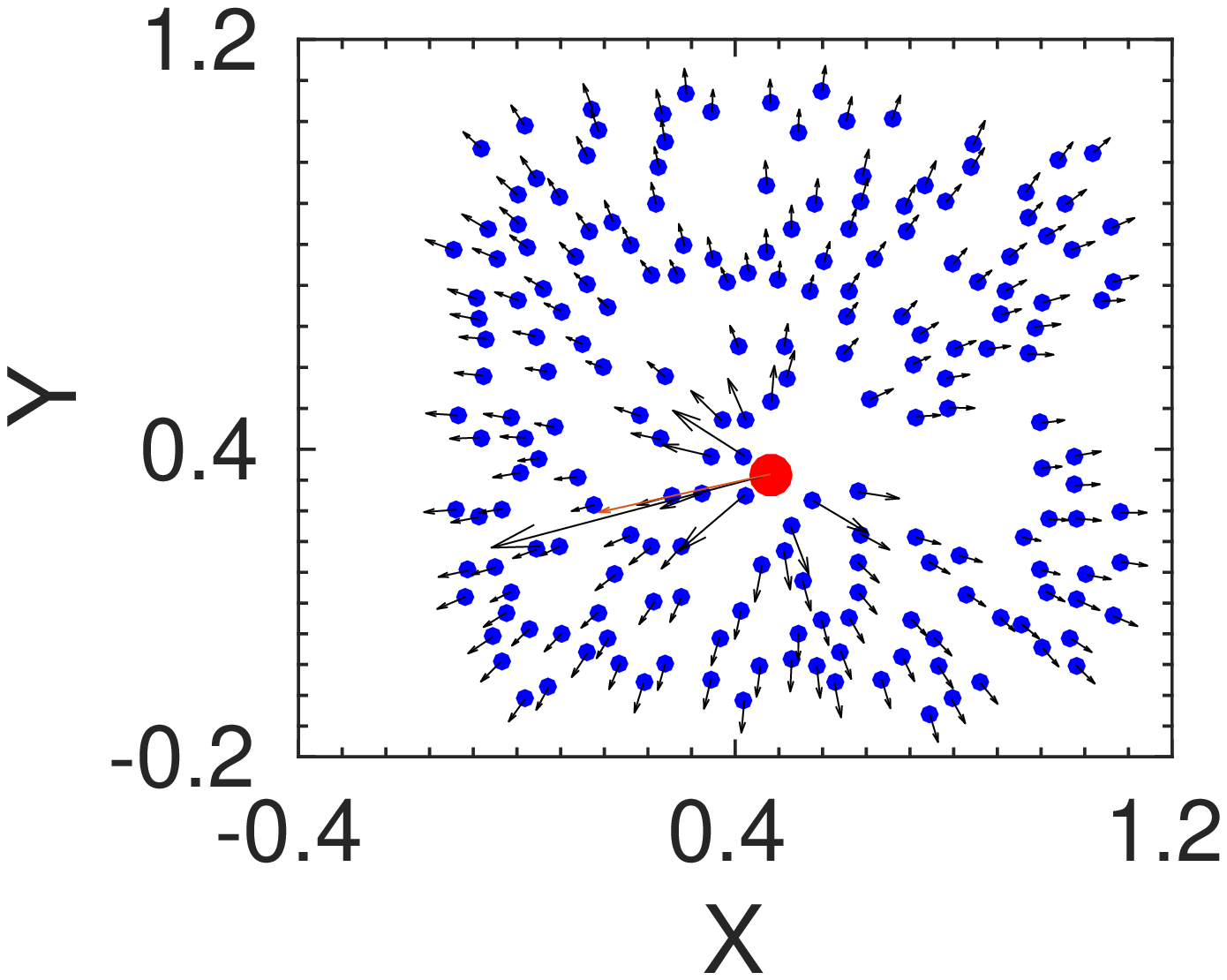}}
	\subfigure[ ]{\label{fig:k}\includegraphics[width=2.8cm,height=2.8cm]{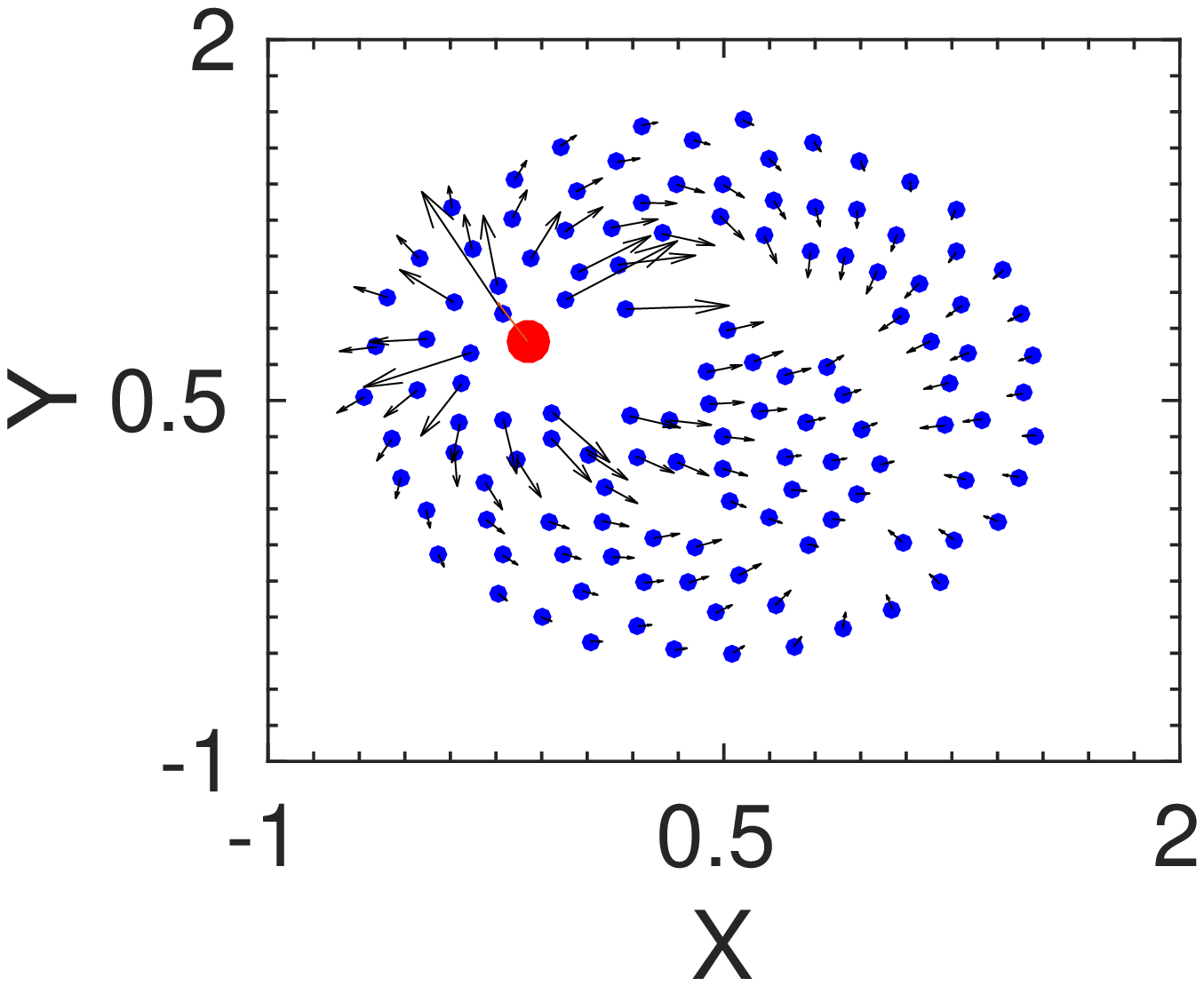}}
	\subfigure[ ]{\label{fig:l}\includegraphics[width=2.8cm,height=2.8cm]{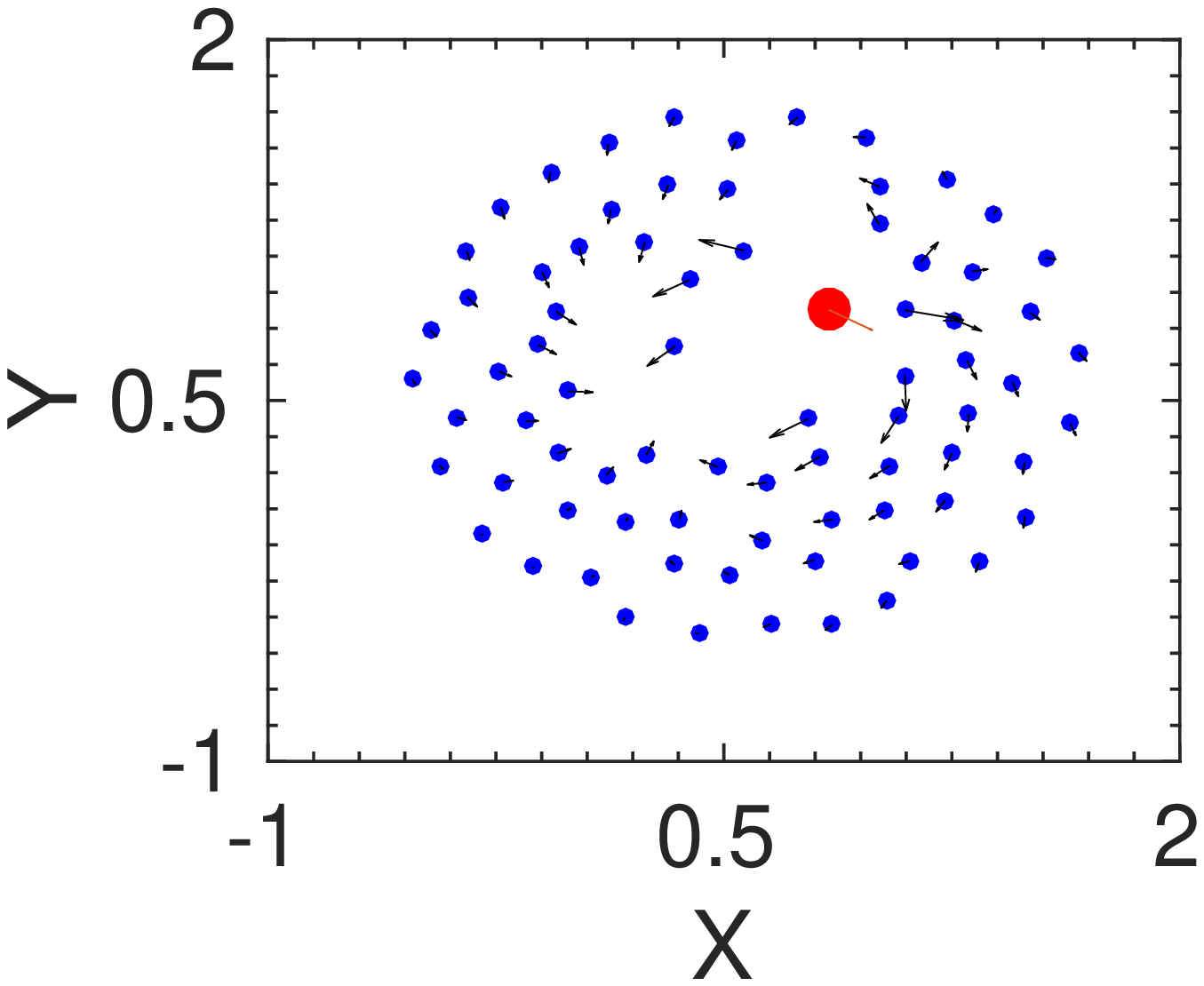}}
	
	\caption{Escape patterns of the prey group (shown by blue dots) under a predator attack (red dot) for different interaction radius, $R_{\rm int}$. (a, b, c) Snapshots of preys and the predator at different simulation time, $T=0.1, 1 , 2$ for $R_{\rm int}=0$, {\it i.e.}, there is no interaction between preys. (d, e, f) Ring formation around the predator at $R_{\rm int}=0.5$; snapshots are at $T=0.1, 10, 1000$. (g, h, i) Splitting up into smaller groups at $R_{\rm int}=1.2$; snapshots are at $T=0.1, 10, 1000$. (j, k, l) For long range interaction, shown here for $R_{\rm int}=2.0$, chasing dynamics of the predator at $T=0.1, 10, 20$. (color online)} 
	\label{phase_plot}
\end{figure}

We have studied the prey-predator dynamics by solving the coupled Eqs. (\ref{preyequation}) and (\ref{predequation}) numerically.  
We have investigated the dynamics for a wide range of parameter values by varying the prey group size $N$, interaction radius $R_{int}$, and also the strength of interactions between preys and the predator. 
Here, we present the dynamics for a case of a `strong' predator. The
`strong' predator signifies that if the prey interacts with all preys to escape from the predator, then the whole group is killed by the predator; on the other hand, in case of a `weak' predator, the whole prey group could easily escape.
The representative parameter values are kept at $\alpha_0=1.0$, $\beta_0=1.0$, $\gamma_0=0.2$, and $\delta_0=2.5$. 
In our simulations, we consider that $N$ preys are initially positioned randomly within a square box of unit area and the predator starts chasing  from just outside the box as illustrated in Fig. \ref{fig:Initialconfig}. 
The kill radius of the predator is taken as $0.01$.
A variety of  escape patterns emerge as we vary the range of interaction radius, $R_{\rm int}$, among preys as shown in different snapshots in Figs \ref{phase_plot}. 
We first consider the scenario, when preys do not interact with each other and every individual just runs away from the predator due to the prey-predator repulsive interaction. 
As seen from Figs. \ref{phase_plot}(a)-(c), the predator hunts down  the randomly moving preys and the whole group get caught over time. 
Now, as we incorporate the prey-prey interactions in the group, we find that for short range of interactions radius,  the whole prey group is also eventually chased down by the predator. However, as the interaction radius increases, different escape patterns emerge.
For example, at interaction radius, $R_{int}=0.5$, as shown in Figs. \ref{phase_plot}(d)-(f), interacting preys form a circle surrounding the predator. Thus, the predator gets confuse which direction to attack and meanwhile the prey group moves away by circling the predator. Such escape routes of ring formation has also been observed in nature for several cases \cite{chenJRSCinterface2014,Partridge1982}. 
Other escaping trajectories also arise by varying the interaction radius further, for example, for $R_{int}=1.2$, the prey group splits into smaller subgroups and migrate away from the predator in small groups as could be seen from Figs. \ref{phase_plot}(g)-(i). Number of subgroups formation not only depends on the interaction radius, but also, on the number of preys and on the initial configurations of the group.
On the other hand, at even larger interaction radius at $R_{\rm int}=2.0$, as shown in  Figs. \ref{phase_plot}(j)-(l), chasing dynamics is observed. The predator is able to catch the prey  and eventually chase down the whole prey group as it has been observed when preys interact with all in the group to avoid the predator.

\begin{figure}[!t]
	\centering  		
     \subfigure[ ]{\label{fig:efficiency}\includegraphics[width=6.5cm,height=5.5cm]{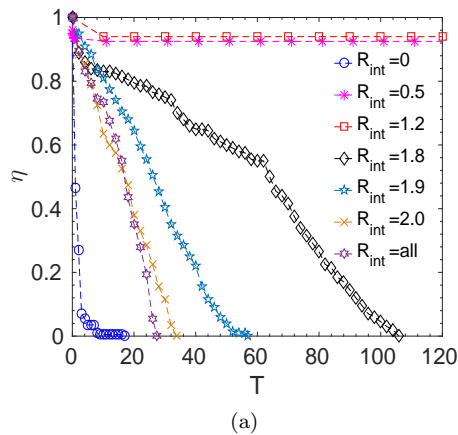}}
     \subfigure[ ]{\label{fig:averagepreyno}\includegraphics[width=6.5cm,height=5.5cm]{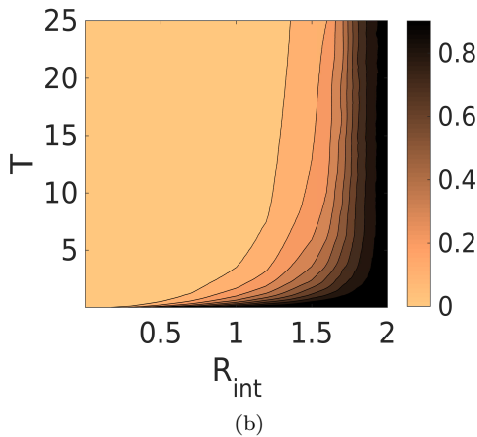}}
	\caption{(a) Survival probability, $\eta$, of the prey group as a function of time, $T$, for different interaction radius, $R_{\rm int}$, and also for the case when all preys interact with each other, $R_{\rm int}= \rm all$, keeping the predator strength constant at $\delta_0=2.5$. (b) Contour plot of the number of interacting preys on an average within an interaction radius, $R_{\rm int}$, normalized by the surviving prey number at that instant as a function of time, $T$, keeping the predator strength, $\delta_0=2.5$. (The color bar indicates the normalized value of the average interacting prey number, {\it e.g.}, the value `$1$' means that each prey on an average is interacting with the entire existing group where as `$0$' means non interacting preys.)}
	\label{fig:4}
\end{figure}

Now, to quantify the survival of preys as a function of interaction radius, we have calculated the survival probability, $\eta$, of the prey group defined by the ratio of number of survived preys, $N_{\rm sur}$, at any instant $T$ to the initial number of preys, {\it i.e.}, 
$\eta=\frac{N_{\rm sur}(T)}{N}$.
It could be seen from  Fig. \ref{fig:efficiency}, as time progresses, for very short range and long range interactions, the whole prey group is killed by the predator. However, in the intermediate range, though initially some preys are caught, but after some time $\eta$ reaches to a steady value which signifies that most of the preys in the group could escape. 
This strong dependence on the interaction range could be understood from the prey-predator dynamics. With no interaction, $R_{int}$=0, or smaller interaction radius, the prey-prey cooperative interaction is not significant and preys move somewhat randomly; hence, due to stronger predator-prey attraction, they are chased down and caught by the predator.
On the other hand, for very large interaction radius among preys is equivalent to interacting with all group members, thus, the whole group  move cohesively; as a result, the predator could easily track the whole prey group and hunt them down. 
Interestingly, in the intermediate range of interaction zone, the initial transient motion show the chasing dynamics by the predator, however, as time progresses, the local interactions of preys eventually establish coordinated movements to confuse the predator by forming a circle or splitting up into sub groups, or by other escape routes to survive in the long run.  
Initially, strong predator-prey attractive force dominates over prey-predator repulsion and prey-prey interaction forces. As time goes on, cooperative interaction force among preys helps to overcome the  attractive force of the predator and thus, preys are survived. 
Such survival strategies of predator  confusion  has also been observed in nature, for example, the hunting behavior of wolves shows that they eventually give up their pursuit after some initial runs (after failed attempts) \cite{mech2010wolves}. 
Our analysis, thus, indicates that a threshold  number of interacting preys is required for cooperative decision making or to confuse the predator. Figure \ref{fig:averagepreyno} shows, how each prey on an average interacts with the number of existing fraction of the population within a given interaction radius while on the chase. As seen, for smaller radius, the interacting prey number is very small, the number increases with increasing $R_{\rm int}$, and after a certain threshold radius, each prey interacts almost with the entire group.
\begin{figure}[!h]
	\centering
	\includegraphics[width=6.0cm,height=5.0cm]{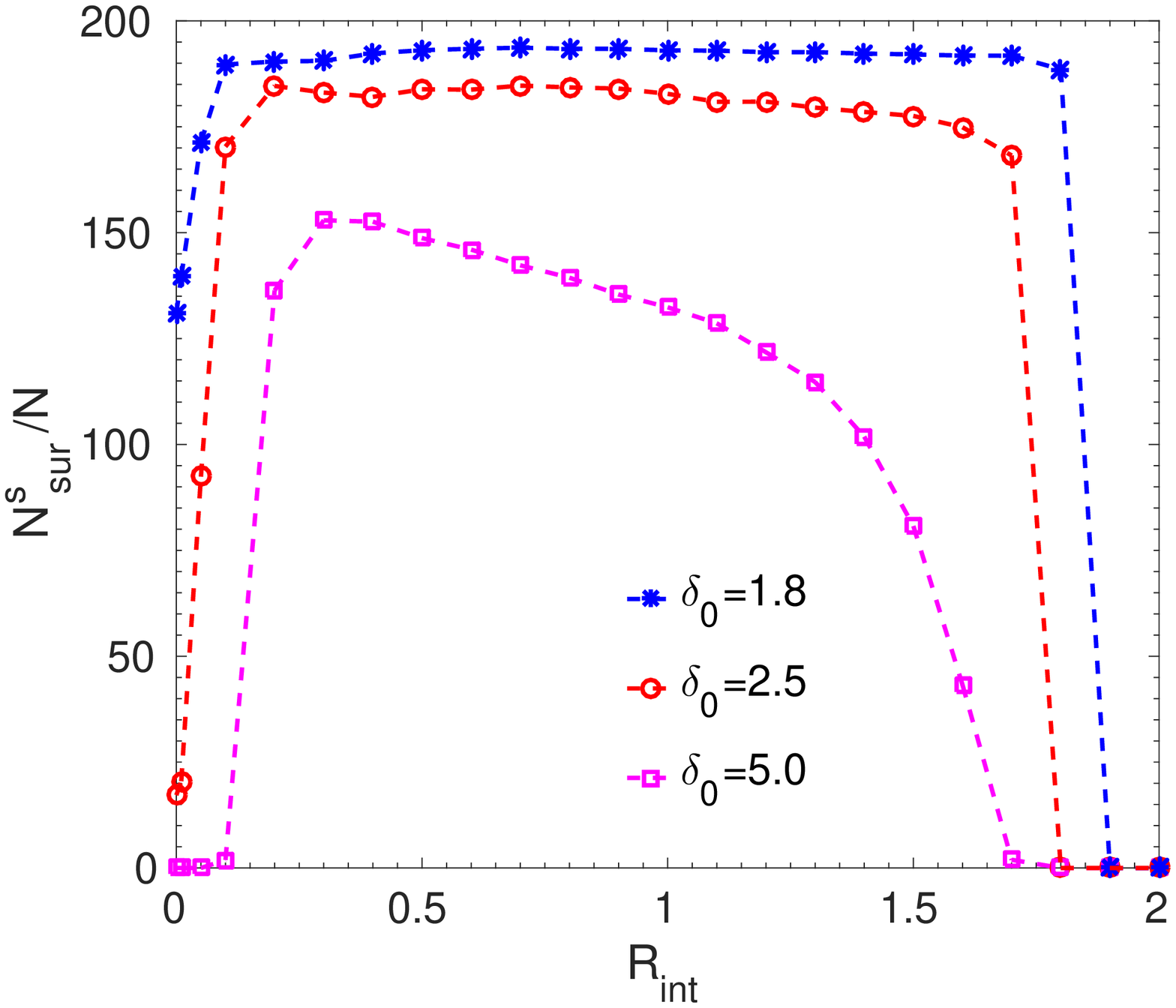}
	\includegraphics[width=6.0cm,height=5.0cm]{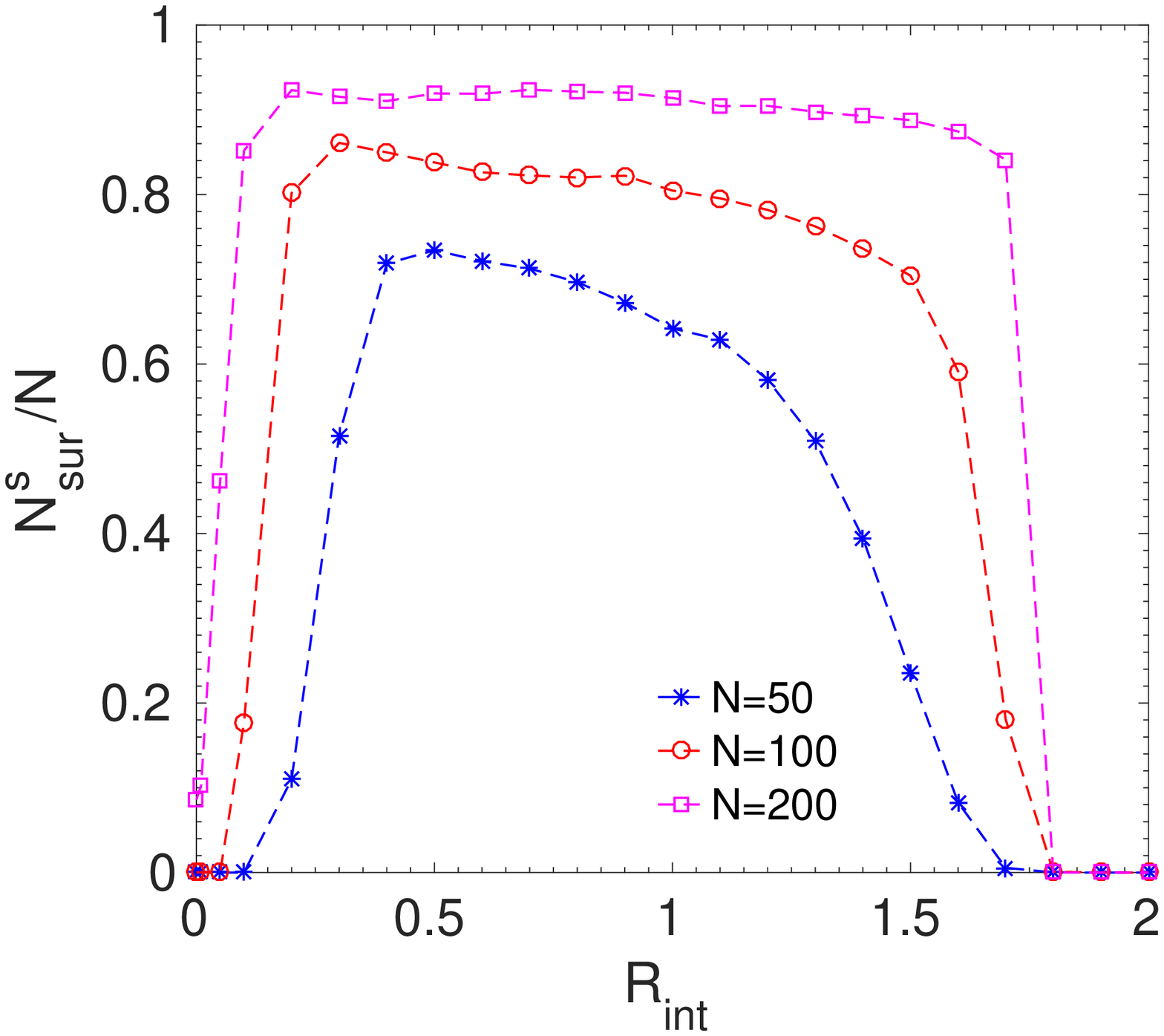}
	\caption{(a) Number of survived prey ($N^{s}_{\rm sur}$)  as a function of interaction radius, $R_{\rm int}$, for different predator strength, $\delta_0=1.8$, $2.5$, and $5.0$ keeping initial group size $N=200$.  (b) Fraction of survived prey number, ($N^{s}_{\rm sur}/N$), as a function of interaction radius, $R_{\rm int}$, for different initial prey group size, $N=50, 100$, and $200$ keeping $\delta_0=2.5$.}
	\label{fig:Nvsrdiffc}
\end{figure}

Further, we  study the number of survived preys  at  steady state, $N^{s}_{\rm sur}$,  as a function of interaction radius as shown in Fig. \ref{fig:Nvsrdiffc}. 
Averaging has been done over two hundred such simulation results.
It can be seen from the plot that very short range and long range interactions  are unfavourable for the prey group survival, however,  
within an intermediate regime the survival of the group is maximum.
Moreover, the survival depends on the prey group size and also on the strength of the predator. In Fig. \ref{fig:Nvsrdiffc}(a), we keep  the initial number of preys at $N=200$,  and vary the strength of the predator, $\delta_0$. It is observed that stronger the predator, lesser the number of survival because the initial catch by the predator is higher for the stronger predator and thus,  the lower threshold value of the interaction radius for the survival of the prey  shifts to the larger $R_{\rm int}$ value as the strength of the predator increases.  Upper threshold value of $R_{\rm int}$ is determined by the range where preys start to interact with almost all existing preys in the group. We now investigate the dynamics by varying the initial prey group size, $N$, while keeping the strength of the predator constant at $\delta_0=2.5$. 
As seen from Fig. \ref{fig:Nvsrdiffc}(b), survival chances of the group increases with increase in the prey group size similar to observations in different field studies \cite{ hirsch2011measuring, cresswell2011predicting}. 
For larger group size, average number of interacting preys  increases within an interaction radius, $R_{\rm int}$. So preys are able to establish coordination in the group to confuse the predator and thus, the survival chances also go up.

\begin{figure}[!h]
	\centering
	\subfigure[ ]{\includegraphics[width=4.0cm,height=3.5cm]{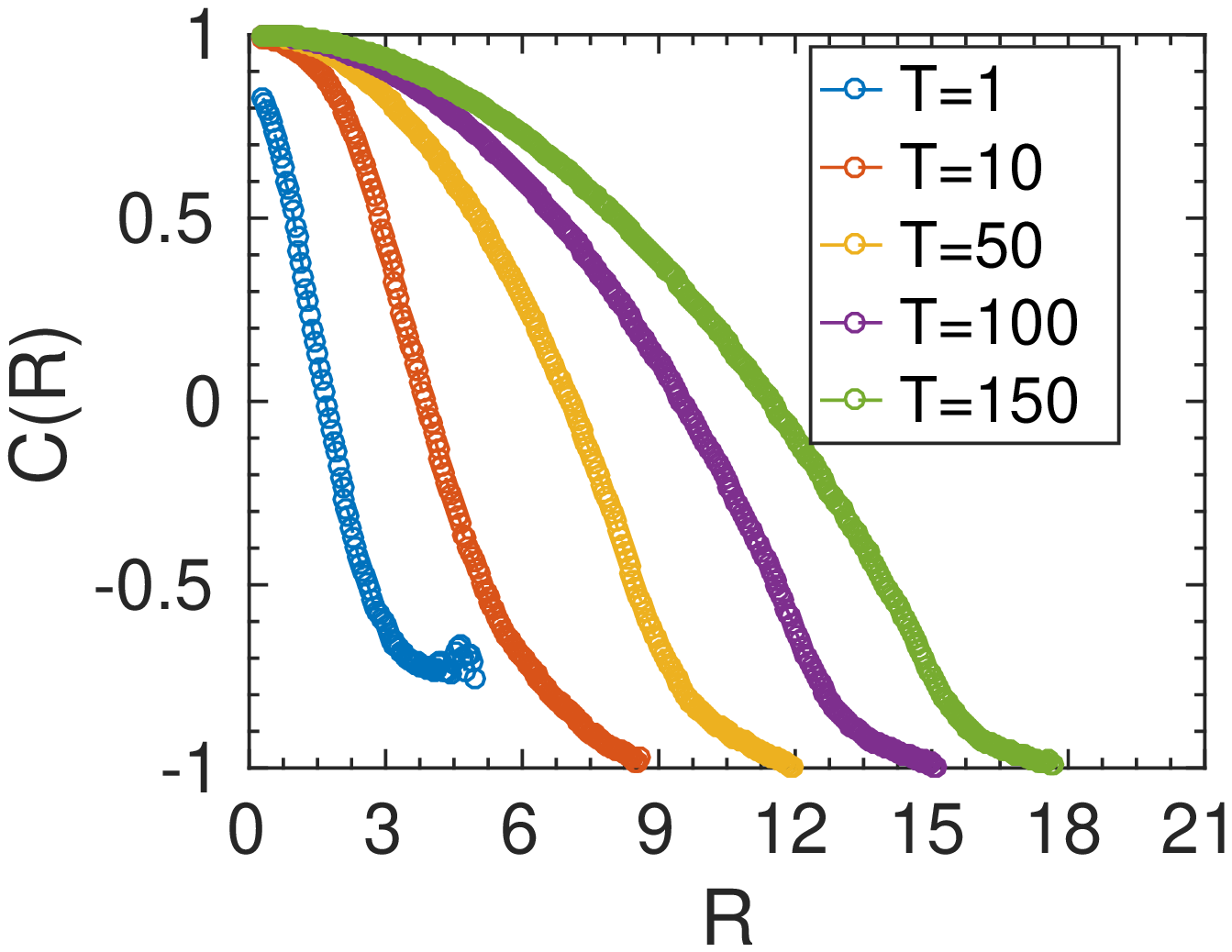}}
	\subfigure[ ]{\includegraphics[width=4.0cm,height=3.5cm]{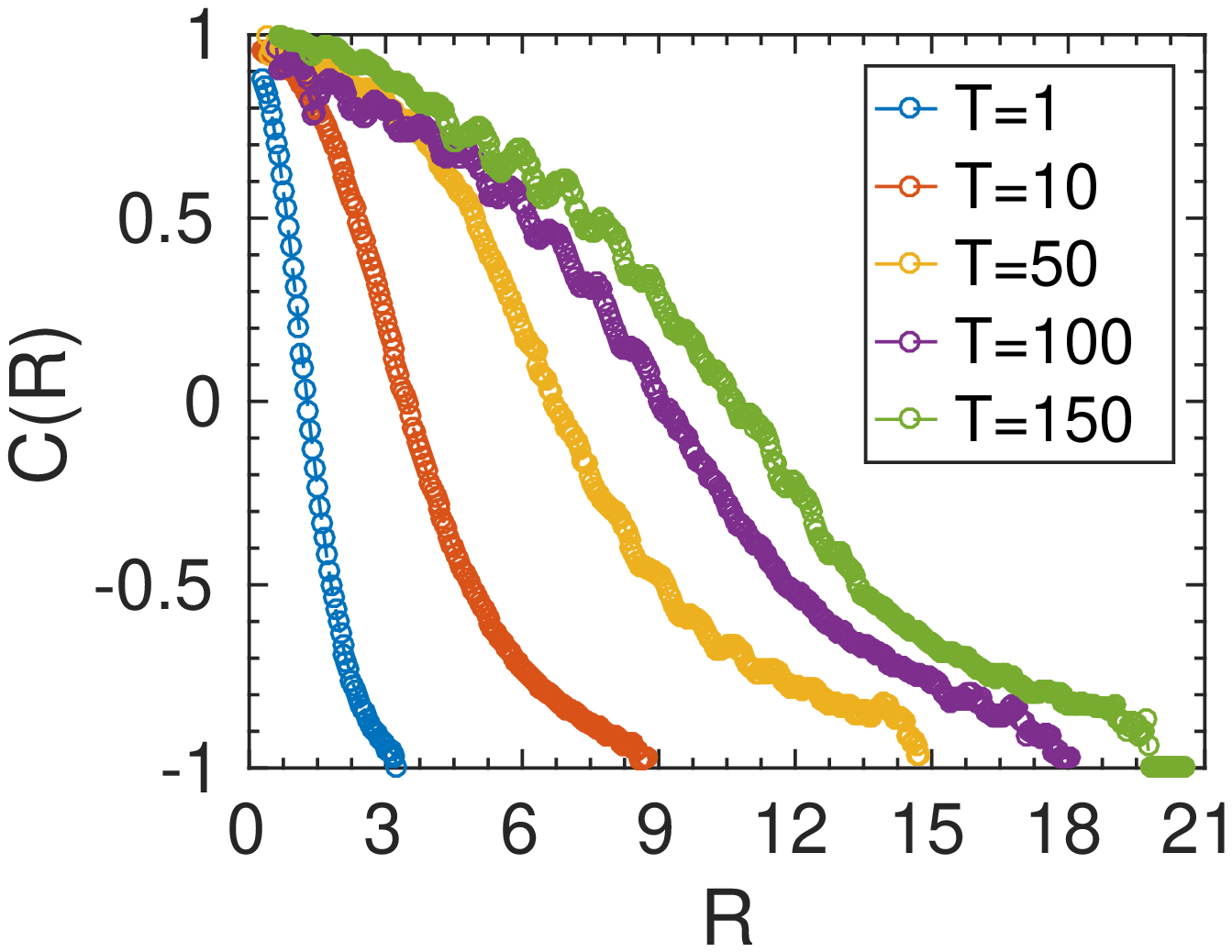}}
	\subfigure[ ]{\includegraphics[width=4.0cm,height=3.5cm]{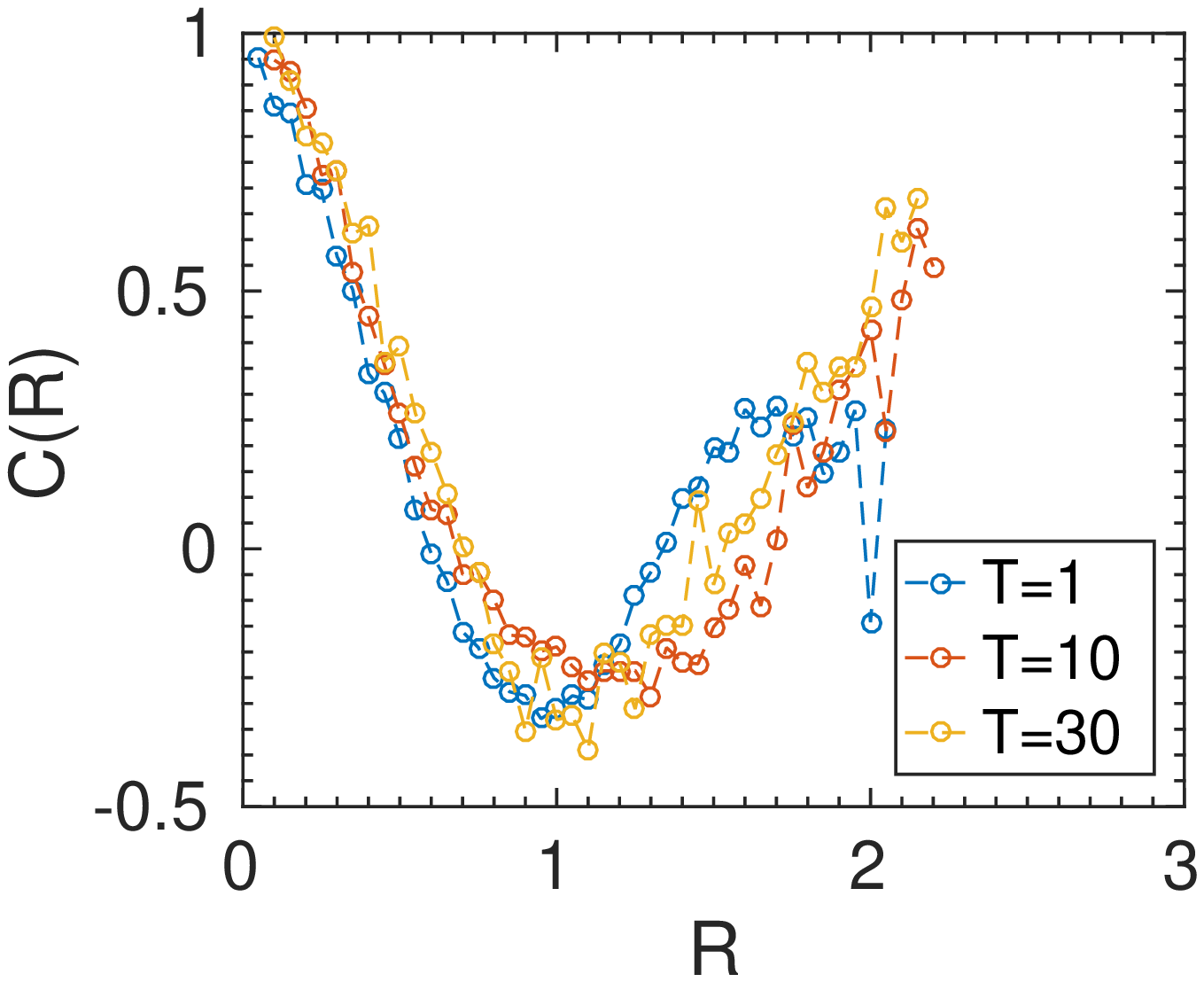}}
	\subfigure[ ]{\includegraphics[width=4.0cm,height=3.5cm]{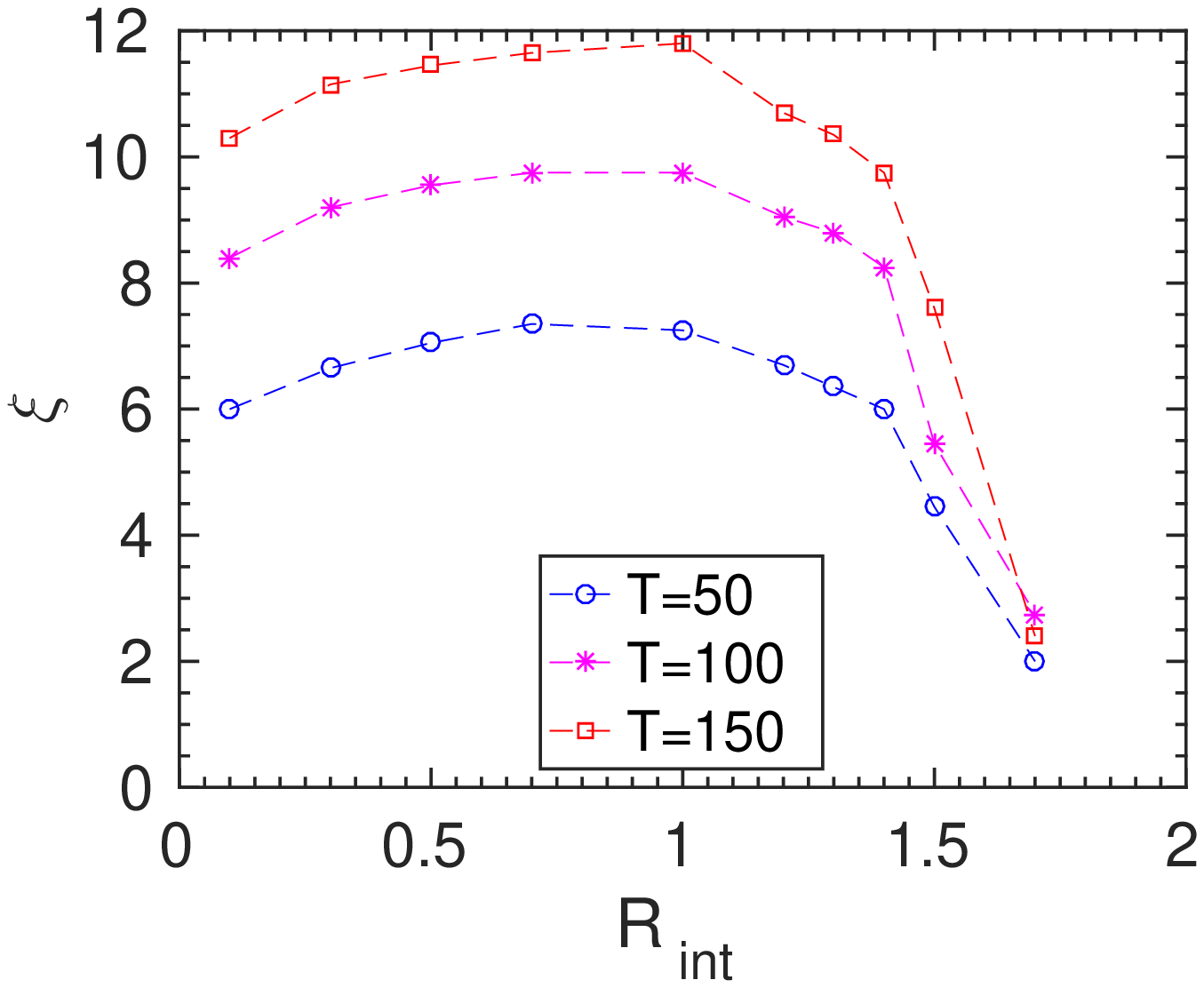}}
	
	\caption{(a),(b),(c) Show spatial correlation in velocity fluctuation, $C(R)$, at different time instances for different interaction radius, $R_{int}=0.5$, $1.2$, and $2.0$ respectively.  (d) Correlation length, $\xi$, as a function of interaction radius, $R_{int}$, for different time. (The predator strength is kept at $\delta_0=2.5$ and the initial prey group size, $N=200$.) }
	\label{fig:spatialcorr}
	
\end{figure}

Moreover, to analyse the collective ordering of the prey group while on escape, we study the spatial correlation in velocity fluctuations 
as described by \cite{attanasi2014collective,attanasi2014finite}, 
\begin{equation}
	C(R)=\frac{\sum_{i,j}{\delta \vec{V}_{i}.\delta \vec{V}_{j} \delta(R - R_{ij})}}{\sum_{i,j}\delta(R - R_{ij})} ;
\end{equation}
where the fluctuation in velocity of the $i$'th prey is defined as 
$\delta \vec{V}_{i}=\vec{V}_{i} - \vec{V}_{\rm av}$, the mean velocity, $\vec{V}_{\rm av}= \frac{1}{N}\sum_i{\vec{V}_{i}}$, and $R_{ij}=|\vec{R}_{i}-\vec{R}_{j}|$ denotes the distance between a pair of preys.
Here, $C(R)$ characterizes how the individual prey behaviour is deviated  from the  average behaviour of the group.
Figure \ref{fig:spatialcorr} shows some representative plots of the spatial correlation, $C(R)$, among  preys for different interaction radius, $R_{\rm int}$, at different time instances. As shown in Figs. \ref{fig:spatialcorr}(a) and (b), within  the
survival regime of the prey-prey interaction, {\it e.g.}, at $R_{\rm int}=0.5$ and  $1.2$, the spatial correlation among  preys increases with time. However, as shown in Fig. \ref{fig:spatialcorr}(c), 
for longer interaction range at $R_{\rm int}=2.0$, representing the non-survival regime, as all preys interact and move cohesively  while chased by the predator, the ordering extends over the entire spatial domain of the group for the whole time period (till the time preys are killed; here, further time instances are not shown as all preys are killed).  We also calculate the  correlation length, $\xi$, for which $C(R=\xi)=0$, to measure the average size of the correlated domain within the prey group in the survival regime. 
Figure \ref{fig:spatialcorr}(d) presents the correlation length, $\xi$, as a function of reaction radius $R_{\rm int}$. As seen, 
$\xi$ increases with increasing $R_{\rm int}$ implying preys become more and more correlated with increase in interaction radius. 
However, with further increase in $R_{\rm int}$, $\xi$ starts 
decreasing as  preys split up into sub groups to escape away from the predator, thus, the correlated domain size  decreases. 

\section*{Discussion}
Here, we show that the range of cooperative interactions in a large group of collectively moving preys is very crucial to strategize their routes of escape under a predator attack. It could be seen from our study, based on a simple theoretical model that accounts for the essential prey-prey and prey-predator interactions, the diverse escape patterns emerges, e.g., ring formation, splitting into subgroups, chasing dynamics etc simply by tuning the  interaction range between preys.  Our study reveals also the survival chances of the group vastly depend on the local range of interacting preys. As shown, selfish run-away of preys without any interaction  is not effective for the survival; similarly, cohesive movements of the entire group  is also unfavourable for the escape. Interestingly, the survival probability is found to be maximum within an intermediate range of interaction radius.  
This work further elucidates the existence of an optimal interaction regime for survival and a certain threshold number of interacting preys to establish the coordinated movements to confuse the predator for escape. Further, under the attack of a weak predator, survival is found to be insensitive to the local interaction range of preys, the whole group could easily escape irrespective of their range of interactions. However, in case of a strong
predator, since  the number of interacting preys are less within a short range, the average prey-prey interaction force is not sufficient to overcome the strong attractive force of the predator. It requires a certain number of interacting preys to confuse the predator and establish the escape routes. 
Moreover, the optimal survival range also depends on the strength of attraction between preys ($\beta_0$ in our model), increasing the strength results  more tightly cohesive prey group and thus becomes more  vulnerable as the predator could easily track and  catch the whole group (for example, keeping the predator strength constant 
at $\delta_0=2.5$, if $\beta_0$ value is increased from $1$ to $2$, then the prey group get killed even at a smaller radius, $R_{\rm int}=1.5$).
The optimal regime is further shown to be sensitive to the prey group size and the strength of the predator as is observed in nature.
Further, our study on spatial correlations in velocity fluctuations in preys show the ordering of the group while on escape. The correlated domain increases with increase in the interaction range among preys, reaches a maximum for a certain radius and then again decreases due to splitting up into smaller subgroups.
Thus, our simple model could shed light into many aspects of natural prey-predator systems. Such theoretical framework could further be extended in understanding of other swarm behaviours of various species, for example, during collective foraging, migratory behaviour of birds or insects to name a few.  Besides, as theoretical modelling and empirical data analysis work hand in hand, more complexity could be incorporated into the model for further quantitative understanding of such conceptual questions in natural scenarios.

\section*{ACKNOWLEDGMENTS}
The authors acknowledge the financial support from Science and Engineering Research Board (SERB), Grant No. SR/FTP/PS-105/2013, Department of Science and Technology (DST), India.

\end{document}